\documentclass[aps,prb,preprint]{revtex4-1}
\usepackage{graphicx}
\usepackage{epstopdf}
\usepackage{tabularx}
\usepackage{graphicx}

\begin{document}
\title{Stable Charged Antiparallel Domain Walls in Hyperferroelectrics}

\author{Shi Liu}
\email{sliu@carnegiescience.edu}
\affiliation{Extreme Materials Initiative, Geophysical Laboratory, Carnegie Institution for Science, Washington, D.C. 20015-1305 USA}
\author{R. E. Cohen}
\email{rcohen@carnegiescience.edu}
\affiliation{Extreme Materials Initiative, Geophysical Laboratory, Carnegie Institution for Science, Washington, D.C. 20015-1305 USA}
\affiliation{Department of Earth- and Environmental Sciences, Ludwig Maximilians Universit\"{a}t, Munich 80333, Germany}

\date{\today}
\begin{abstract}{
Charge-neutral 180$^\circ$ domain walls that separate domains of antiparallel polarization directions are common structural topological defects in ferroelectrics. In normal ferroelectrics, charged 180$^\circ$ domain walls running perpendicular to the polarization directions are highly energetically unfavorable because of the depolarization field and are difficult to stabilize.  We explore both neutral and charged 180$^\circ$ domain walls in hyperferroelectrics, a class of proper ferroelectrics with persistent polarization in the presence of a depolarization field, using density functional theory. We obtain zero temperature equilibrium structures of head-to-head and tail-to-tail walls in recently discovered $ABC$-type hexagonal hyperferroelectrics. Charged domain walls can also be stabilized in canonical ferroelectrics represented by LiNbO$_3$ without any dopants, defects or mechanical clamping. First-principles electronic structure calculations show that charged domain walls can reduce and even close the band gap of host materials and support quasi-two-dimensional electron(hole) gas with enhanced electrical conductivity.  
}
\end{abstract}
\maketitle

{\bf Introductions}

A ferroelectric domain wall (DW) is an interface separating domains of different polarization directions in ferroelectrics.~\cite{Pramanick12p243} Because DWs can move in response to external stimuli such as electric field and mechanical stress, their presence can substantially affect the electro-mechanical and electro-thermal coupling properties of ferroelectrics.~\cite{Zhang94p454,Taylor97p1973,Xu01p1336,Xu14p3120,Karthik11p024102,Karthik12p167601,Zubko16p524} DWs possessing symmetries different from their parent bulk materials can also exhibit unique properties that do not exist in the bulk. One notable example is the electrical conducting DWs in ferroelectric semiconductors such as BiFeO$_3$,~\cite{Seidel09p229,Maksymovych11p1906,Farokhipoor11p127601} Pb(Zr,Ti)O$_3$,~\cite{Guyonnet11p5377,Tselev16p11630} and LiNbO$_3$.~\cite{Schrder12p3936} Ekhard Salje {\em et\ al.~}also demonstrated the polarity and ferroelectircity in ferroelastic DWs in SrTiO$_3$~\cite{Salje13p247603} and CaTiO$_3$.~\cite{Yokota14p144109} The ability to tune and control the conductive DW in an insulating medium via external electrical or stress field provides a new playground for designing new nanoelectronics.~\cite{Catalan12p119} Recent experiments also highlighted the critical role of DWs for enhancing photovoltaic current in BiFeO$_3$~\cite{Bhatnagar13p2835,Yang17p43070} and facilitating electron-hole separations in CH$_3$NH$_3$PbI$_3$.~\cite{Liu15p693}

The DW type is defined based on the angle formed between adjacent polarization vectors.~\cite{Pramanick12p243} The 180$^\circ$ wall separates domains with antiparallel polarization directions. Depending on the orientation of the wall relative to the direction of polarization inside adjacent domains, the 180$^\circ$ DW can have three configurations: a neutral wall running parallel to the polarization direction, a head-to-head (HH) configuration with polarization directed towards the wall, and a tail-to-tail (TT) configuration with polarization directed from the wall. The polarization discontinuity at the HH (TT) wall will give rise to positive (negative) bound charges, which if unscreened will create a large depolarization electric field, leading to unfavorable electrostatic energy and destabilization of ferroelectricity.~\cite{Gureev11p184104,Wu06p020103}  The neutral 180$^\circ$ DW has been observed and prepared in a wide range of ferroelectrics experimentally and has been the subject of numerous theoretical studies with density functional theory (DFT) and molecular dynamics.~\cite{Meyer02p104111,Poykko99p2830,
Wojde14p247603,Liu17p094102,Lee09p060102,Lee10p014104} However, strongly charged 180$^\circ$ walls are much less common in ferroelectrics: they are rarely formed naturally due to their high formation energies. In normal ferroelectrics, their stabilization requires some extrinsic mechanism (e.g., dopant and defects) to compensate the polarization-induced bound charges.~\cite{Gureev11p184104} Recently periodic charged 90$^\circ$ DWs, stabilized by both the free-carrier compensation at the wall and the elastic compatibility of adjacent ferroelastic domains, were prepared in BaTiO$_3$ thin films by carefully controlling the poling history and mechanical boundary conditions.~\cite{Sluka13p1808,Bednyakov15p15819} The free-carrier compensated charged DWs can support quasi-two-dimensional electron gas (q2DEG) resulting in steady metallic-type conductivity,~\cite{Vul73p29,Gureev11p184104,Sluka13p1808} and therefore are potentially useful for technology applications when combined with the electric field-tunable DW mobility. Wu and Vanderbilt studied HH and TT 180$^\circ$ DWs in PbTiO$_3$, where the wall was stabilized by the intentional insertion of charged-impurity layers (Sc$^{3+}$ at HH and Nb$^{5+}$ at TT).~\cite{Wu06p020103} Here we explore the intrinsic properties of defect/dopant-free charged DWs.

Recent seminar work on hexagonal $ABC$ semiconducting ferroelectrics using DFT calculations revealed hyperferroelectricity~\cite{Garrity14p127601}  characterized by persistent polarization even at ideal open circuit condition.~\cite{Fu14p164104} Normal ferroelectrics in their high-symmetry nonpolar phases have at least one unstable transverse optic (TO) mode and stable longitudinal optic (LO) modes. The origin of ferroelectricity in normal ferroelectrics comes from the delicate balance between the long-range Coulomb interaction that favors the polar phase and the short-range repulsion that favors the nonpolar phase.~\cite{Cohen92p136} Hyperferroelectrics, on the other hand, in the high-symmetry phase possess both TO and LO-mode instability.~\cite{Garrity14p127601} The imaginary LO phonon frequency is attributed to the small LO-TO splitting resulting from the large optical dielectric constants and small Born effective charges. It was recently pointed out that the fundamental driving mechanism for hyperferroelectricity is the short-range interaction, which already favors the symmetry-broken polar state.~\cite{Li16p34085} LiNbO$_3$, a prototypical normal ferroelectric, is suggested to be a hyperferroelectric as well, with the short-range instability of Li also contributing to the ferroelectricity and being robust against electric boundary conditions. The persistent polarization in hyperferroelectrics in the presence of a depolarization field hints at easier formation of strongly charged DWs with HH and TT configurations.~\cite{Garrity14p127601}  In this work, we study both charged and uncharged 180$^\circ$ DWs in several known hyperferroelectrics with DFT calculations. 

{\bf Computational Methods} 

We studied 180$^\circ$ DWs in $ABC$-type ferroelectrics, LiBeSb, LiBeBi, LiZnAs, NaZnSb, and KMgBi,  and Li$B$O$_3$-type ferroelectrics represented by LiNbO$_3$. These materials were suggested to be hyperferroelectrics.~\cite{Garrity14p127601,Li16p34085} 
The structure of $ABC$ ferroelectric is a hexagonal variant of the half-Heusler structure with a polar space group of $P6_3mc$: the hexagonal unit cell has six atoms with $B$ and $C$ atoms forming buckling honeycomb layers separated by layers of ``stuffing" $A$ atoms.~\cite{Bennett12p167602} Ferroelectric LiNbO$_3$ belongs to the $R3c$ space group and the hexagonal unit cell has 30 atoms with the spontaneous polarization aligned along the $c$ axis ($z$ direction). The neutral 180$^\circ$ DW is modeled with a $Na\times1a\times1c$ supercell ($a$ and $c$ are short-axis and long-axis lattice constants of a hexagonal unit cell, respectively), where the unit cells are stacked in the $x$ direction ($a$ axis) and $N/2$ unit cells have polarization aligned along $+z$, and $N/2$ unit cells have polarization aligned along $-z$.
The charged 180$^\circ$ DW is modeled with a $1a\times1a\times Nc$ supercell with the unit cells stacking along the $z$ direction and polarization changing from $+z$ to $-z$. We choose $N = 8$ for $ABC$ ferroelectrics and $N=4$ for Li$B$O$_3$ ferroelectrics. Because of periodic boundary conditions, the supercells contain two neural walls or one TT wall and one HH wall. The DW energy ($E_{\rm DW}$) is calculated with $E_{\rm DW} = \frac{1}{2S}\left(E_{\rm supercell} - E_{\rm SD}\right)$, where $S$ is the DW area, $E_{\rm supercell}$ is the energy of the supercell with two DWs, and $E_{\rm SD}$ is the energy of the fully-relaxed single-domain supercell of the same number and stacking of unit cells. All calculations are carried out within the local-density approximation (LDA) using Quantum Espresso~\cite{Giannozzi09p395502} with a $1\times4\times4$ Monkhorst-Pack $k$-point grid for supercells with neutral walls, and a $4\times4\times1$ $k$-point grid for supercells containing charged walls. A force convergence threshold of 5.0$\times 10^{-5}$ Ry/Bohr,  an energy convergence threshold of 1.0$\times 10^{-4}$ Ry, and Mazrzari-Vanderbilt smearing of 1 mRy are used to fully relax the dimensions of the supercell and atomic positions. We used ultrasoft pseudopotentials from the Garrity, Bennett, Rabe, Vanderbilt (GBRV) high-throughput pseudopotential set~\cite{Garrity14p446} and a plane-wave cutoff of 50 Ry and charge density cutoff of 250 Ryd. 

{\bf Results and Discussion}

\begin{figure}[t]
\centering
\includegraphics[scale=1]{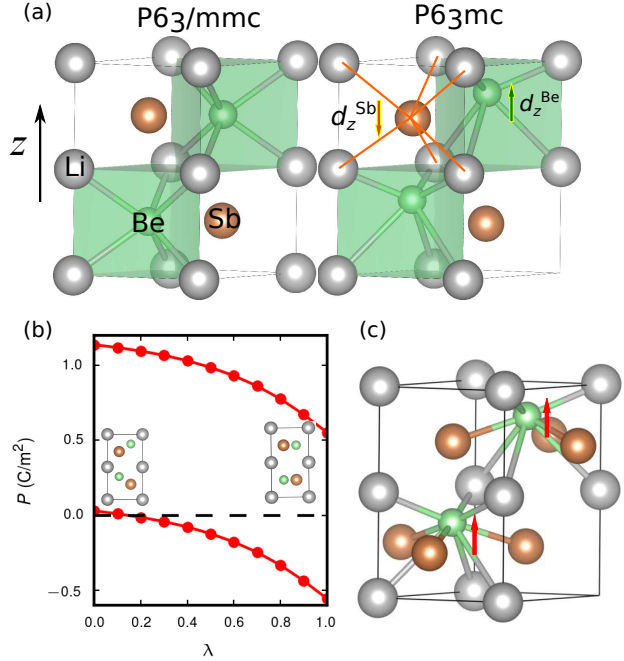}\\
 \caption{ (a) Crystal structures of LiBeSb in the high-symmetry $P6_3/mmc$ phase and the polar $P6_3mc$ phase. The $c$ axis is aligned along the $z$ direction. The Be is displaced along $+z$ with respect to the center of the Li$_6$ cage whereas the Sb is displaced along $-z$. (b) Calculation of the effective polarization with the Berry-phase approach. The structure is changed adiabatically from $P6_3mc$ phase to $P6_3/mmc$ phase. (c) Structural motif for the calculation of local polarization at Be. Each Be is shared by three Sb atoms and six Li atoms. 
  }
  \label{Unitcell}
 \end{figure}

{\em Spontaneous polarization in LiBeSb } We start by examining the spontaneous polarization of LiBeSb in the polar $P6_3mc$ structure, which relates to the nonpolar $P6_3/mmc$ structure through the buckling of the BeSb honeycomb layers. The fully relaxed $P6_3mc$ structure has $a=4.083$~\AA~and $c=6.620$~\AA, consistent with previous studies with LDA.~\cite{Bennett12p167602,Garrity14p127601} Structurally, the polarization along the $c$ axis (aligned along the $z$ direction) in $P6_3mc$ structure results from the displacement of Be and Sb away from the center of the Li$_6$ octahedron (FIG.~\ref{Unitcell}a). The Be and Sb atomic displacements ($d_z^{\rm Be}$ and $d_z^{\rm Sb}$) are calculated to be 0.68~\AA~ and -0.09~\AA, respectively. We first estimate the polarization with the Berry-phase approach by tracking the change in Berry phase while adiabatically transforming the structure from $P6_3mc$ phase to $P6_3/mmc$ phase (FIG.~\ref{Unitcell}b). The total effective polarization (defined relative to a centrosymmetric reference) is found to be $P_z = 0.58$ C/m$^2$, agreeing well with previously reported value of 0.59 C/m$^2$.~\cite{Bennett12p167602,Garrity14p127601} It is noted that the formal polarization (the raw result of Berry-phase calculation)~\cite{Rabe07Book} in the nonpolar $P6_3/mmc$ structure is exactly half a quantum of polarization. The formal polarization does not necessarily vanish in a centrosymmetric materials such as III-III perovskite (e.g., LaAlO$_3$) where the individual layers (LaO and AlO$_2$) are not charge neutral.~\cite{Vanderbilt93p4442,Stengel09p241103} This is also the case for $ABC$ ferroelectrics where the $A$ and $BC$ layers have formal charges of $\pm e$.   

We can also estimate $P_z$ by summing the product of atomic displacements and Born effective charges (BECs), $P_z=\frac{e}{\Omega}\left(d_z^{\rm Se}Z_{zz}^{\rm Se} + d_z^{\rm Be}Z_{zz}^{\rm Be}\right)$, where $Z_{zz}$ is the Born effective charge along $z$ and $\Omega$ is the volume per formula unit (one half of the unit cell volume).  The underlying assumption of this approach is that BECs have similar values in polar and refernece nonpolar structures. It works well for normal ferroelectrics such as PbTiO$_3$ and BaTiO$_3$ where BECs are not sensitive to structural details,~\cite{Zhong94p3618} and gives polarization value close to that obtained with the Berry-phase approach.~\cite{Neaton05p014113} However, the BECs of Be and Sb undergo substantial changes when the structure transforms from polar $P6_3mc$ phase to nonpolar $P6_3/mmc$ phase, as BeSb layers moving from an $sp^3$ bonding to $sp^2$ bonding enviroment. Using the BECs in $P6_3mc$ phase ($Z_{zz}^{\rm Be}  = 0.561$ and $Z_{zz}^{\rm Se} =-1.836$), we obtain $P_z = 0.185$ C/m$^2$, whereas using the BECs in $P6_3/mmc$ phase ($Z_{zz}^{\rm Be}  = 4.656$ and $Z_{zz}^{\rm Se} =-5.847$), we obtain $P_z = 1.244$ C/m$^2$, neither reproducing the right polarization value. Finally, using the mean BECs of $P6_3mc$ and $P6_3/mmc$ phases, $P_z$ is found to be 0.72 C/m$^2$, roughly agreeing with the Berry phase approach. This behavior was also noticed in a recent DFT study on LiBeSb.~\cite{Dai16p034103}

{\em Energetics }  We obtain both neutral and charged 180$^\circ$ DWs in all studied hyperferroelectrics in the absence of any dopants or mechanical clamping. Charged 180$^\circ$ DWs of HH and TT configurations are highly unstable in prototypical ferroelectrics such as PbTiO$_3$: even within zero-Kelvin DFT calculations, the supercell containing HH and TT walls will eventually transform to a single domain during structural optimization process. In hyperferroelectrics, the fully-optimized charged 180$^\circ$ walls maintained the HH or TT configurations. Table 1 reports the optimized structures and estimated energetics for both neutral and charged walls. The energetics of neutral 180$^\circ$ DWs in $ABC$ ferroelectrics are comparable to that in PbTiO$_3$ (102 mJ/m$^2$).~\cite{Liu17p094102}  In all cases, the charged domain walls have much higher energy compared to their neutral counterparts. Recent DFT investigations suggested that KMgBi is a hyperferroelectric topological insulator which supports both persistent polarization and metallic topological surface states.~\cite{Sante16p076401} Interestingly, the DW energy in KMgBi is also the lowest among all studied hyperferroelectrics. Exploring how ferroelectric DWs may interact with topological surface states~\cite{Liu16p1663} will be a useful future research topic.
 
{\em Atomistic DW structure in LiBeSb } We provide a detailed analysis of the 180$^\circ$ DW structures in $ABC$ hyperferroelectrics by taking LiBeSb as an example. To give a quantitative description, we compute the local atomic displacement for Be and Sb and also the local polarization centered at Li, Be and Sb ($\mathbf{P}^{\rm Li}$, $\mathbf{P}^{\rm Be}$ and $\mathbf{P}^{\rm Se}$). 
The local polarization at Be is calculated with  
\begin{equation}
\mathbf{P}^{\rm Be} = \frac{e}{\Omega}\left(\frac{1}{3}\mathbf{Z}^{\rm Sb}\sum_{i=1}^{3}\mathbf{r}^{{\rm Sb}}_i  +\mathbf{Z}^{\rm Be}\mathbf{r}^{{\rm Be}} + \frac{1}{6}\mathbf{Z}^{\rm Li}\sum_{i=1}^{6}\mathbf{r}^{{\rm Li}}_i \right)
\end{equation}
where $\mathbf{Z}$ is the BEC tensor and $\mathbf{r}$ is the atomic position. 
Equation 1 is essentially the polarization resulting from the Be-centered local dipole moment formed by the Be and its nearest three Sb atoms and six Li atoms (FIG.~\ref{Unitcell}c). Similarly, $\mathbf{P}^{\rm Se}$ and  $\mathbf{P}^{\rm Li}$ are defined as
\begin{equation}
\mathbf{P}^{\rm Sb} = \frac{e}{\Omega}\left(\frac{1}{6}\mathbf{Z}^{\rm Li}\sum_{i=1}^{6}\mathbf{r}^{{\rm Li}}_i + \frac{1}{3}\mathbf{Z}^{\rm Be}\sum_{i=1}^{3}\mathbf{r}^{{\rm Be}}_i  +\mathbf{Z}^{\rm Sb}\mathbf{r}^{{\rm Sb}} \right)
\end{equation}
and 
\begin{equation}
\mathbf{P}^{\rm Li} = \frac{e}{\Omega}\left(\mathbf{Z}^{\rm Li}\mathbf{r}^{{\rm Li}} +\frac{1}{6}\mathbf{Z}^{\rm Be}\sum_{i=1}^{6}\mathbf{r}^{{\rm Be}}_i  + \frac{1}{6}\mathbf{Z}^{\rm Sb}\sum_{i=1}^{6}\mathbf{r}^{{\rm Sb}}_i \right)
\end{equation}
We used the mean BECs of $P6_3mc$ and $P6_3/mmc$ phases when calculating the local polarization. 

\begin{figure}[t]
\centering
\includegraphics[scale=0.5]{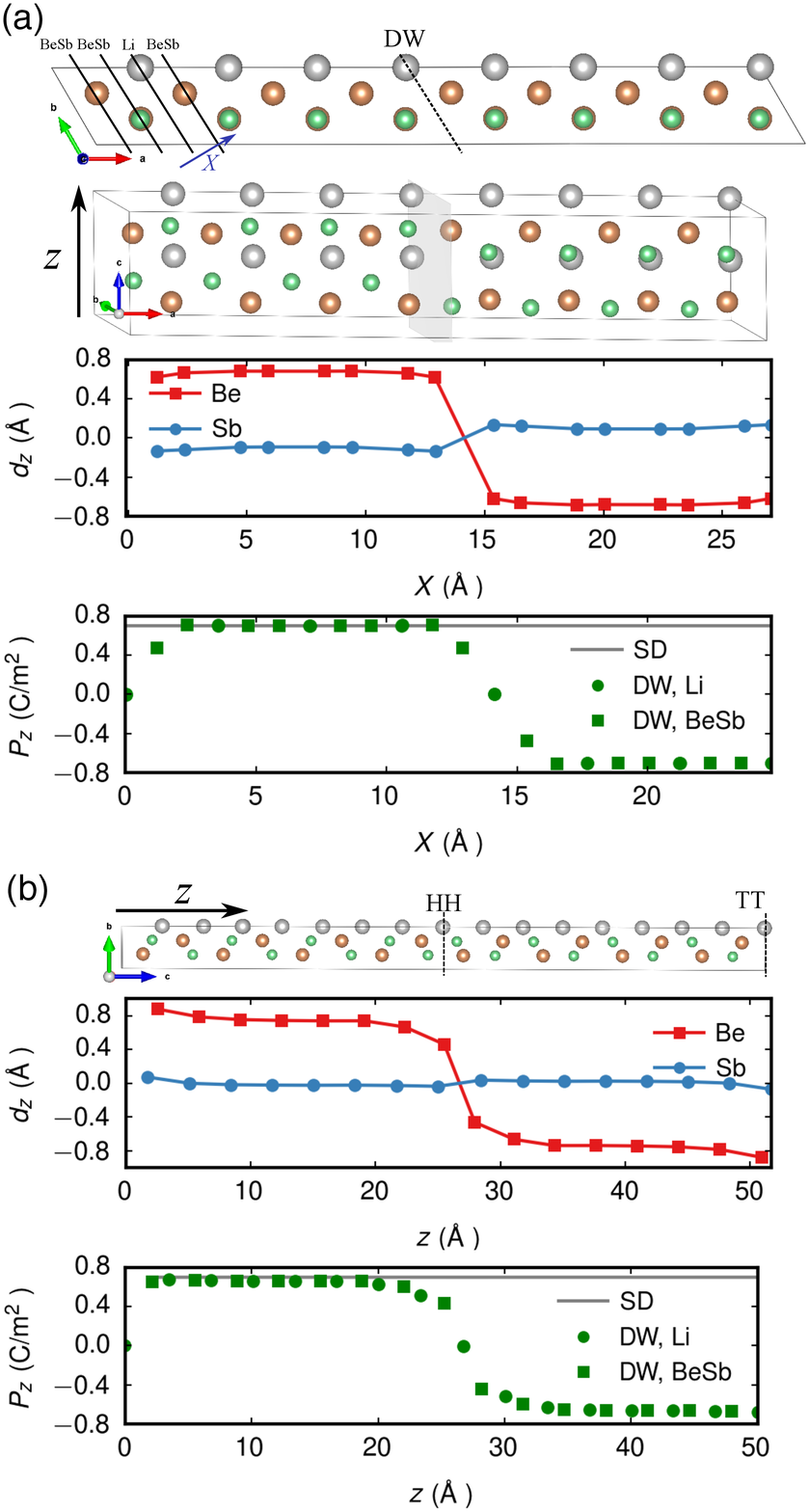}\\
 \caption{ Atomic displacements ($d_z$) and polarization profiles ($P_z$) for supercells with (a) neutral and  (b) charged 180$^\circ$ domain walls (DW) in LiBeSb.  The polarization profile for the single domain (SD) supercell is also plotted as the reference. }
  \label{LiBeSb_DW}
 \end{figure}

The neutral 180$^\circ$ wall lies parallel to the $ab$ plane (FIG.~\ref{LiBeSb_DW}a). We calculate the layer-resolved polarization for lattice planes (alternating planes consisted of Li and  BeSb) stacked along the $X$ direction (normal to the $ab$ plane). The neutral wall centers at the Li layer characterized by zero local polarization. This is similar to Pb-centered  180$^\circ$ DWs in PbTiO$_3$ with nearly zero local polarization at the PbO lattice plane. We fit the polarization profile to $P_0{\rm tanh} [{(z-z_0)}/{\xi_{\rm DW}}]$, where $z_0$ and 2$\xi_{\rm DW}$ correspond to the center and the width of the DW~\cite{Meyer02p104111} and find 2$\xi_{\rm DW}=2.86$~\AA, which is about one unit cell along the $a$ axis. For the supercell containing charged DWs (FIG.~\ref{LiBeSb_DW}b), the magnitude of the polarization in the internal domain-like region is similar to that in a single domain, suggesting a nearly perfect screening of the depolarization field. Charged DWs also center at Li planes with $P_z^{\rm Li} =0$~C/m$^2$. The HH and TT walls show one subtle difference: Be atoms near the HH wall have $d_z$ smaller than that in domain-like region whereas they change the direction abruptly when crossing the TT wall without reducing the magnitude. For DW width we find a value of 2$\xi_{\rm DW}$ = 6.4~\AA~for the HH wall and 2$\xi_{\rm DW}$ = 1.4~\AA~for the TT wall, showing the TT wall is much sharper. This is likely due to the positive Li atoms at the TT wall that help to compensate the negative boundary charges. 

\begin{figure}[t]
\centering
\includegraphics[scale=0.5]{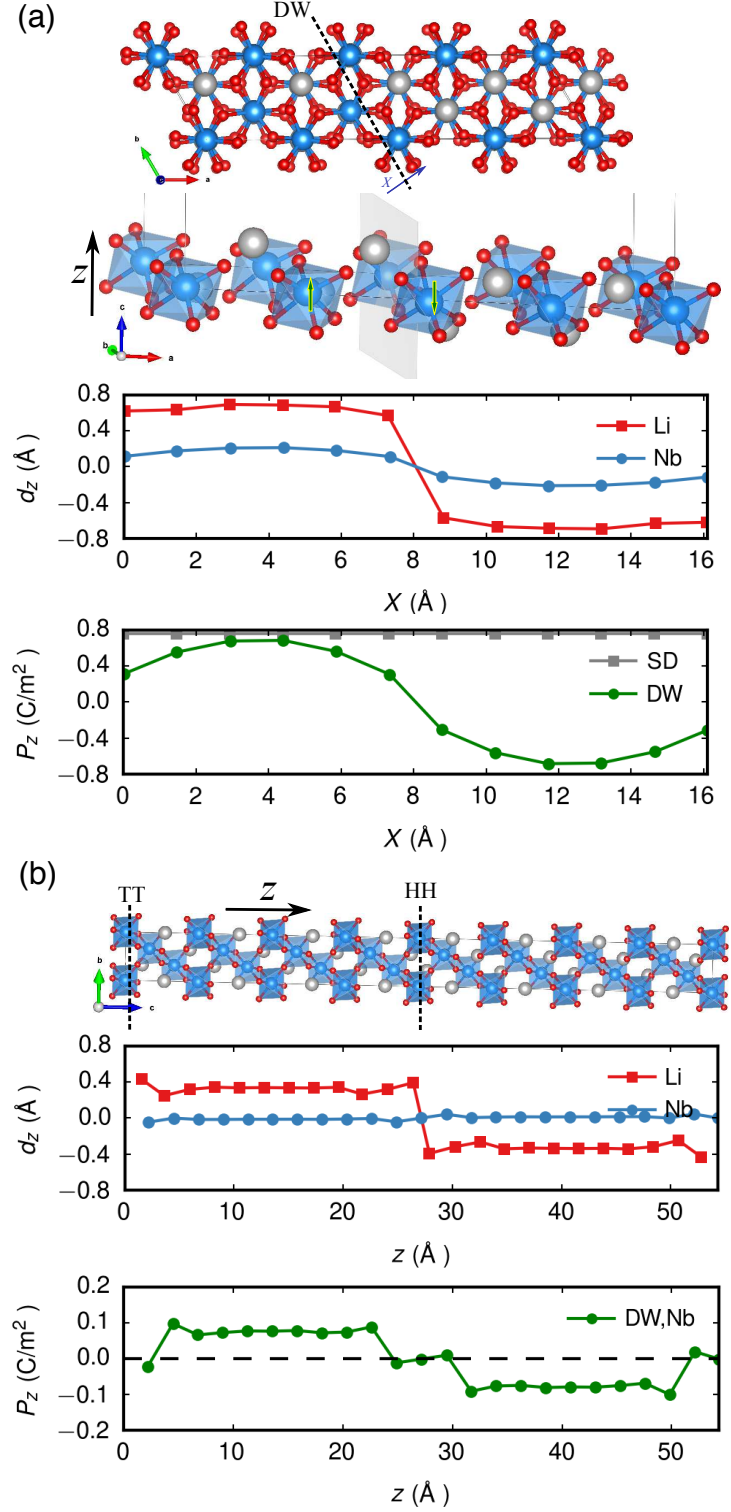}\\
 \caption{Atomic displacements ($d_z$) and polarization profiles ($P_z$) for supercells with (a) neutral and  (b) charged 180$^\circ$ domain walls (DW) in LiNbO$_3$. }
  \label{LNO_DW}
 \end{figure}

{\em Atomistic DW structure in LiNbO$_3$} The neutral 180$^\circ$ walls in LiNbO$_3$ have been studied with different computational methods such as DFT and molecular dynamics, and Ginzburg-Landau-Devonshire (GLD) theory.~\cite{Lee09p060102,Lee10p014104,Scrymgeour05p184110,Ye17014105} There are two crystallographically different DWs with X wall lying parallel to a mixed anion-cation plane and a Y wall running parallel to alternating planes consisted of only cations and only anions.~\cite{Gopalan07p449} As the main focus of this work is on charged DWs, we here only studied the neutral X wall.
To reveal the DW structure in LiNbO$_3$, we consider LiNbO$_3$ as a distorted perovskite and calculate the atomic displacements of Li and Nb atoms with respect to the center of their surrounding oxygen cages (O$_{12}$ for Li and O$_6$ for Nb). The local polarization at Nb ($\mathbf{P}^{\rm Nb}$) is defined as 
\begin{equation}
\mathbf{P}^{\rm Nb} = \frac{e}{\Omega}\left(\frac{1}{8}\mathbf{Z}^{\rm Li}\sum_{i=1}^{8}\mathbf{r}^{{\rm Li}}_i  +\mathbf{Z}^{\rm Nb}\mathbf{r}^{{\rm Nb}} + \frac{1}{6}\mathbf{Z}^{\rm O}\sum_{i=1}^{6}\mathbf{r}^{{\rm O}}_i \right)
\end{equation}
Our DFT calculations show that the X wall lies halfway between the ion planes (FIG.~\ref{LNO_DW}a), consistent with previous results.~\cite{Lee09p060102,Lee10p014104,Ye17014105} The polarization and atomic displacements change the direction across the X wall without becoming zero.

Although LiNbO$_3$ has been intensely studied with first-principles methods,~\cite{Inbar96p1193,Lee09p060102,Lee10p014104,Scrymgeour05p184110,Ye17014105} our work for the first time reveals the equilibrium structure of fully-relaxed HH and TT walls (FIG.~\ref{LNO_DW}b) in this canonical ferroelectric. We find that Nb atoms are little displaced locally across the whole supercell, whereas the Li atoms still have relatively large atomic displacements ($\approx 0.34$~\AA), indicating the polarization should primarily come from the distortion of Li atoms. This agrees with recent understanding of hyperferroelectric instability in Li$B$O$_3$ due to short-range interactions of Li that are less sensitive to electric boundary conditions. The internal domain-like regions remain to be polar albeit adopting polarization ($P^{\rm Domain}\approx 0.076$~C/m$^2$) much smaller than the bulk polarization ($\approx 0.76$~C/m$^2$). The polarization reduction is driven by the unscreened depolarization field arising from the bound charges at DWs. The calculated domain polarization ($P^{\rm Domain}$) in the presence of charged DWs actually agrees quite well with the reported value of polarization under zero displacement field condition ($P_{D=0}=0.08$~C/m$^2$) for LiNbO$_3$.~\cite{Li16p34085}

\begin{figure}[t]
\centering
\includegraphics[scale=1.5]{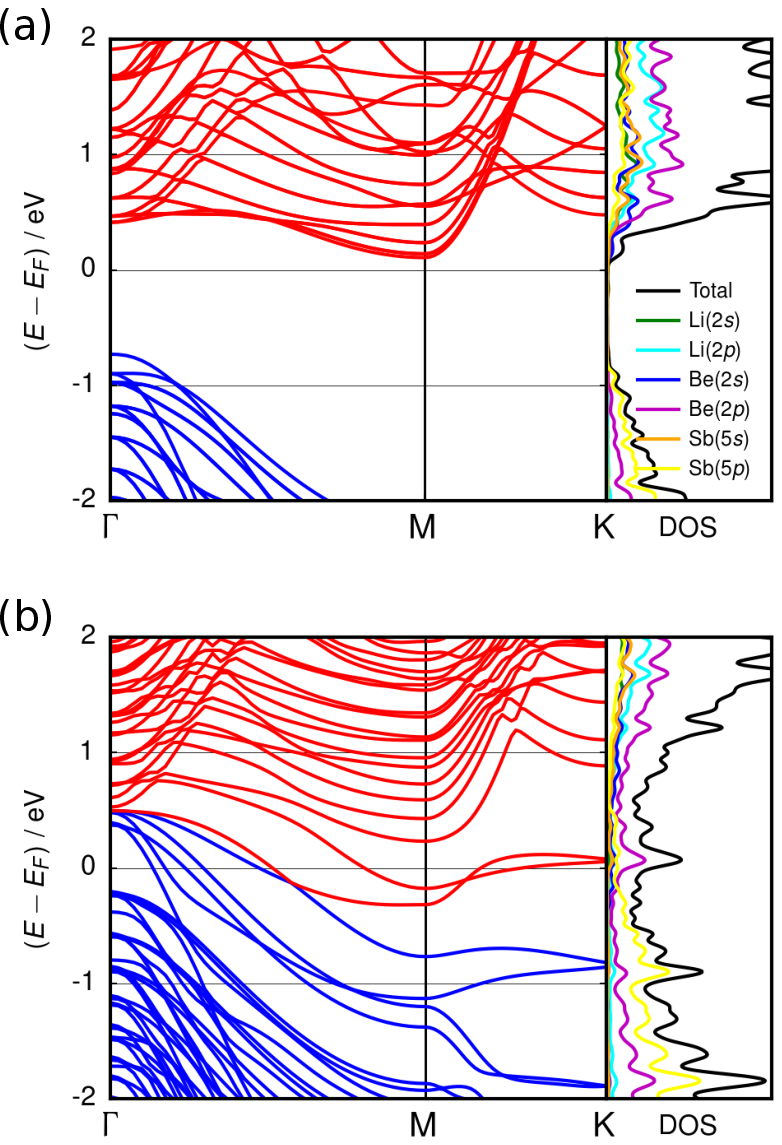}\\
 \caption{ Band structures and orbital-resolved density of states for (a) a single-domain supercell and (b) a supercell with charged 180$^\circ$ domain walls in LiBeSb. }
  \label{LiBeSb_BandStructure}
 \end{figure}

{\em Electronics structure } 
The presence of charged DWs can significantly influence the electronic properties of host materials.~\cite{Sluka12p748,Liu15p693} Unless the bound charges are fully compensated, the depolarization field due to the imperfect compensation will result in an electrostatic potential step across the domain sandwiched by the charged DWs,  shifting the energy of band edge states. At the positive HH wall, the conduction band will be pushed downward the Fermi energy with a tendency to create free electrons to screen the positive bound charge. Similarly, the valence band will approach the Fermi level to provide free holes to compensate the negative bound charge at the TT wall. For a large enough potential step (that scales with the distance between the TT and HH walls), the band gap can be closed and the structure will become metallic. We compare the band structures and orbital-resolved density of states (DOS) for supercells of LiBeSb with and without charged DWs (FIG.~\ref{LiBeSb_BandStructure}). The Brillouin zone $k$-points are increased to $16\times16\times2$ for DOS calculations. The band gap for the single-domain LiBeSb is 0.84 eV using LDA, with $p$ orbitals of Be and Sb atoms contributing to the band-edge valence bands and $s$ and $p$ orbitals of Li, Be and Sb atoms all contributing to the conductions bands (FIG.~\ref{LiBeSb_BandStructure}a).

\begin{figure}[t]
\centering
\includegraphics[scale=1.5]{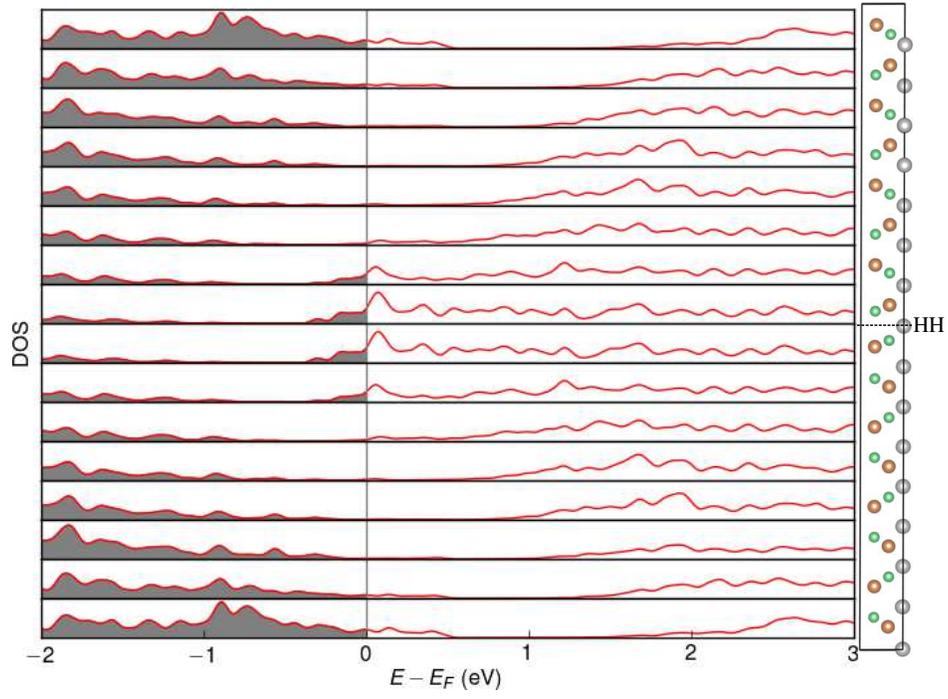}\\
 \caption{Layer-resolved density of states for a supercell with charged 180$^\circ$ domain walls in LiBeSb. }
  \label{LiBeSb_DOS}
 \end{figure}

\begin{figure}[t]
\centering
\includegraphics[scale=1.0]{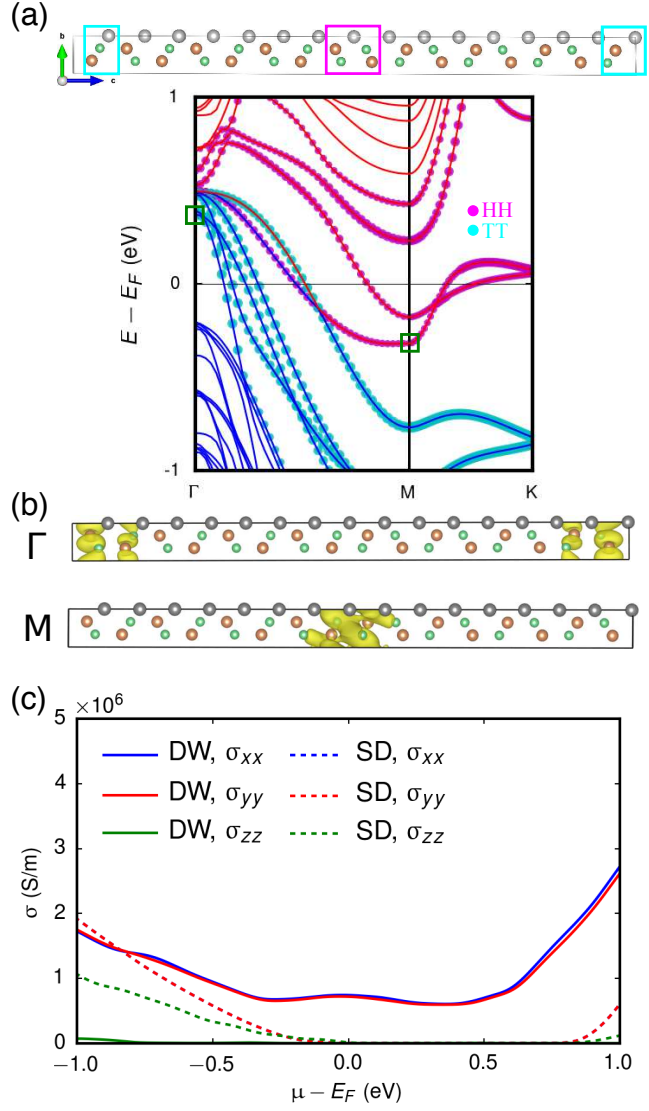}\\
 \caption{(a) Atomic orbital-resolved band structure for a supercell with charged HH and TT walls. The magenta and cyan circles represent the contribution of states from atoms within the HH and TT wall, respectively. The size of the circle scales with the contribution. (b) Charge density plots projected on $\rm \Gamma$ and M points of two bands highlighted in (a). (c) Electrical conductivity versus the Fermi levels for structures with and without charged domain walls .}
  \label{LiBeSb_BandStructure2}
 \end{figure}

Because of the relatively small band gap (at the LDA level), the $1a\times 1a\times 8c$ LiBeSb supercell with charged walls separated by 26.8~\AA~(4$c$) is already metallic with states near the Fermi level primarily consisted of Be $2p$ and Sb $5p$ orbitals (FIG.~\ref{LiBeSb_BandStructure}b). The layer-resolved DOS reveals more details of the electronic structure in real space (FIG.~\ref{LiBeSb_DOS}). The conduction band minimum (CBM) is located at the HH wall and the valence band maximum (VBM) is located at the TT wall, both crossing the Fermi level ($E_F$) and providing free carriers. In contrast, the layers between charged DWs remain mostly insulating. This demonstrates the metallicity comes from the charged DWs. The width of the conducting layers is $\approx$13.4~\AA~(about two unit cells along the $c$-axis), indicating the presence of quasi-two-dimensional electron gas and hole gas (q2DEG and q2DHG). We further project the band structure onto orthogonalized atomic wave functions, and the weight of atomic wave functions from atoms at HH and TT walls are then evaluated separately for nine bands near the Fermi energy (FIG.~\ref{LiBeSb_BandStructure2}a). It is evident that conduction bands near the Fermi level are dominated by atomic orbitals of atoms within the HH wall and the valence bands providing free holes are mainly consisted of states of atoms at the TT wall, consistent with the layer-resolved DOS analysis.  The charge density plots in FIG.~\ref{LiBeSb_BandStructure2}b  show the spacial extension of the 2DEG (2DHG).  We also calculate the electrical conductivity as a function of Fermi level (assuming a constant carrier scattering relaxation time of 10 fs) using Boltzmann transport equation with BoltzTrap package.~\cite{Madsen06p67}  For transport calculations, a $32\times32\times4$ $k$-point sampling is used for electronic structure calculations. As shown in FIG.~\ref{LiBeSb_BandStructure2}c, charged DWs significantly enhance the electrical conductivity within the plane of DW ($\sigma_{xx}$ and $\sigma_{yy}$), whereas the conductivity normal to the DW remains low ($\sigma_{zz}$). 

\begin{figure}[t]
\centering
\includegraphics[scale=1.5]{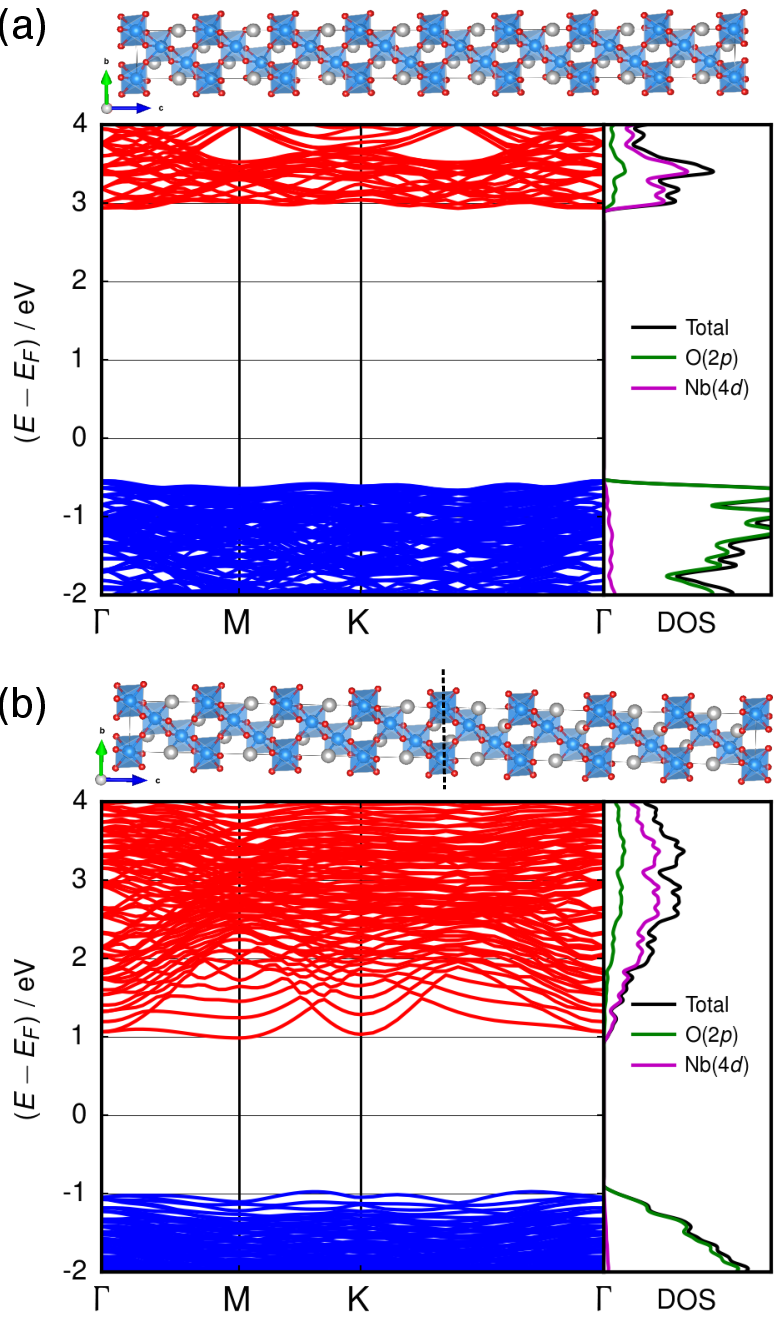}\\
 \caption{Band structures and orbital-resolved density of states for (a) a single-domain supercell and (b) a supercell with charged 180$^\circ$ domain walls in LiNbO$_3$. }
  \label{LNO_BandStructure}
 \end{figure}

As shown in Figure~\ref{LNO_BandStructure}, the LDA band gap for single-domain LiNbO$_3$ is 3.48 eV and it reduces to 1.96 eV after introducing charged DWs separated by 27.2~\AA~(4$c$) (half length of the  $1a\times 1a\times 4c$ supercell along the $c$ axis). With increasing distance ($L$) between charged DWs, the electrostatic potential step will eventually exceed the band gap of LiNbO$_3$ and drive the DWs mettallic, similar to the case of LiBeSb. The $L$-dependence of the band gap roughly follows $E_g = E_g^{\rm SD} - 2P_{D=0}L / \varepsilon_c $, where $E_g^{\rm SD}$ = 3.48 eV is theoretical bulk band gap, $P_{D=0}$ is the polarization at zero displacement field and $\varepsilon_c$ is the dielectric constant along the $c$ axis. Taking $P_D=0.08$~C/m$^2$,  $\varepsilon_c=30$ and $L=27.2$~\AA, we obtain $E_g=1.84$~eV, agreeing reasonably well with DFT values. The critical distance $L$ that closes the gap is estimated to be 58~\AA.

{\bf Conclusion}

The structure and energetics of neutral and charged 180$^\circ$ domain walls in several hyperferroelectrics have been studied by density functional theory.  The fully-relaxed charged domain walls in $ABC$ hyperferroelectrics are surprisingly narrow and their widths are comparable to neutral walls. Taking LiBeSb as an example, we find that the polarization bound charges at charged walls are nearly perfect screened by the free carriers with the head-to-head wall supporting quasi-two-dimensional electron gas and the tail-to-tail wall supporting quasi-two-dimensional hole gas. In LiNbO$_3$, we also obtain strongly charged 180$^\circ$ walls separating bulk-like regions of smaller polarization. Compared to Nb distortion, the Li distortion is more robust against the depolarization field. Because of the large band gap of bulk LiNbO$_3$, a large distance between charged walls is required to close the band gap.  Understanding the stability of  strongly charged walls in hyperferroelectrics at finite temperatures and their mobility in response to electric field will be useful future research topics. 

\newpage
\noindent {\bf Acknowledgements} This work is partly supported by US Office of Naval Research Grants N00014-12-1-1038 and N00014-14-1-0561. 
SL and REC are supported by the Carnegie Institution for Science. REC is also supported by the
European Research Council Advanced Grant ToMCaT. Computational support was provided by theß US DOD through a Challenge Grant from the HPCMO. SL acknowledges Dr. Kevin F. Garrity for sharing structural files of $ABC$ ferroelectrics.  
%

\newpage

\begin{table}
\centering
\caption{Optimized Supercell Dimensions and Energetics of Neutral and Charged DWs (nDWs and cDWs) in Hyperferroelectrics.}
\begin{tabular}{l|l|l|l|l|l|l|c}
\hline 
  & $l_x$ & $l_y$ & $l_z$ & $\alpha$ & $\beta$ & $\gamma$ & $E_{\rm DW}$ (mJ/m$^2$) \\ 
\hline 
LiBeSb,nDW &32.717 &4.103 &6.629 &90.00 &89.99& 120.10 &192 \\ 
\hline 
LiBeSb,cDW &4.090 &4.080 &53.638 &89.96 &90.00 &120.07 &1232  \\ 
\hline 
LiBeBi,nDW &33.642& 4.229& 6.835& 90.00& 89.99& 120.19& 169\\
\hline
LiBeBi,cDW &4.175 &4.169 &56.128 &90.00 &91.99 &119.95& 535\\
\hline 
LiZnAs,nDW &32.572 &4.093& 6.585& 90.00 &90.00 &120.18& 174\\
\hline 
LiZnAs,cDW &4.089 &4.126& 51.541& 90.01& 89.99& 119.72& 1332\\
\hline 
NaZnSb,nDW &35.991 &4.520 &7.239 &90.00& 90.00 &120.16 &84 \\
\hline 
NaZnSb,cDW &4.517 &4.538 &57.595& 90.00 &90.00 &120.07 &567\\
\hline
KMgBi,nDW &40.594 &5.077 &7.891 &90.00 &90.00 &120.02 &14\\
\hline
KMgBi,cDW &5.089 &5.089 &62.765 &90.00 &90.00 &120.00 &369\\
\hline
LiNbO$_3$,nDW &  20.301 &5.076& 13.661 &90.00 &90.01 &120.00& 140\\
\hline
LiNbO$_3$,cDW &  5.110 &5.110 &54.359 &90.00 &90.00 &120.00 &1044\\
\hline 
\hline
\end{tabular} 
\end{table}

\newpage
 
 \newpage
 
  \newpage
 
  \newpage
 
   \newpage
  
  \newpage
 
   \newpage

\end{document}